# A platform for lightweight deployment of IoT applications based on a Function-as-a-Service model

S. Sansó, C. Guerrero*, I. Lera, and C. Juiz, *Senior Member, IEEE*

*Abstract*— This paper presents a platform to facilitate the deployement of applications in Internet of Things (IoT) devices. The platform allows to the programmers to use a Function-as-a-Service programming paradigm that are managed and configured in a Platform-as-a-Service web tool. The tool also allows to establish interoperability between the functions of the applications. The proposed platform obtained faster and easier deployments of the applications and the resource usages of the IoT devices also were lower in relation to a deployment process based in containers of Docker.

*Keywords*— Internet of Things, Function-as-a-Service, Application orchestration, Cloud computing.

## I. INTRODUCCIÓN

DURANTE los últimos años, se ha producido un incremento importante del número de aplicaciones desarrolladas para ecosistemas basados en el Internet de las Cosas o IoT. En este tipo de aplicaciones, se considera que cualquier dispositivo, por pequeño que sea, es capaz de conectarse a Internet y de monitorizar o controlar elementos físicos. Estos dispositivos recogen, procesan y comparten datos. Uno de los principales problemas de estos entornos es la heterogeneidad de los dispositivos IoT, que provoca que el despliegue de estos entornos sea un trabajo laborioso y costoso [1].

Con la adopción de las arquitecturas IoT en entornos tan diversos como los industriales (*Industrial IoT*, IIoT), inteligentes (*Smart cities*), de transporte, o domésticos (*Smart homes*), se ha incrementado la necesidad de ofrecer plataformas para un despliegue rápido y efectivo de aplicaciones IoT. Existen esfuerzos previos de propuestas de plataformas con este fin, pero estas están, por un lado, basadas en modelos de desarrollo que requieren de herramientas muy específicas y que no están disponibles para todos los tipos de dispositivos IoT y, por otro lado, requieren de altos requisitos tanto de espacio como de rendimiento en dichos dispositivos, hecho que también eleva la utilización de los recursos de red cuando se lleva a cabo el despliegue de las aplicaciones.

Estos inconvenientes han de ser solucionados mediante la adopción de un modelo de desarrollo que sea capaz de reducir el tamaño y los requerimientos de las aplicaciones ejecutadas en los dispositivos IoT. Las aplicaciones IoT que se ejecutan en los dispositivos suelen ser aplicaciones muy simples que se encargan únicamente de leer datos (monitorizar sensores) o escribir información (accionar actuadores) con ciertas reglas o condiciones. El modelo de desarrollo que se ha dado en llamar *Function-as-a-Service* (FaaS) satisface estos requerimientos, ya que las aplicaciones son desarrolladas como funciones que se ejecutan en cualquier elemento de proceso disponible [2].

Adicionalmente, el aumento de los requisitos, tanto de almacenamiento de los datos monitorizados por los sensores, como de su procesado para la toma automática de decisiones, ha provocado que sea necesario disponer de elementos con mayores capacidades de proceso que los propios dispositivos IoT. Los servicios basados en la nube, o *Cloud Computing*, cubren estas necesidades. Gracias a la interrelación del binomio IoT y Cloud, se pueden desarrollar aplicaciones que interaccionen con dispositivos reales, gracias al IoT, y que al mismo tiempo tengan capacidades de almacenamiento y procesado ilimitadas, gracias al Cloud. No es nueva la idea de soportar aplicaciones que interaccionan con el mundo real mediante el uso de servicios en la nube y ya ha sido previamente utilizada en otros campos, como por ejemplo la computación móvil o la robótica [3].

Con la incorporación de las tecnologías Cloud, se pueden definir plataformas para aplicaciones IoT que faciliten y simplifiquen el proceso de despliegue de las mismas.

En este artículo presentamos una plataforma para el despliegue de aplicaciones IoT desarrolladas usando un modelo FaaS, y cuya orquestación se realiza mediante servicios en cloud que utilizan un modelo de *Platform-as-a-Service* (PaaS).

## II. ESTADO DEL ARTE

Como se ha dicho anteriormente, la heterogeneidad de los sistemas IoT es aún un campo de investigación con problemas abiertos. Por ejemplo, Yacchimera y Palau [4] propusieron un Gateway para solucionar los problemas de heterogeneidad de los dispositivos IoT, mediante la defición de un protocolo flexible que transforma los datos heterogéneos en un formato común.

La definición de plataformas de despliegue de aplicaciones en entornos similares a las arquitecturas IoT ya se ha investigado anteriormente. En la mayoría de casos, estas soluciones se basan en modelos que generan un mayor consumo de recursos en los dispositivos donde se despliegan y que la disponibilidad de dichas herramientas es más limitada que otras tecnologías más tradicionales. Es por ejemplo el caso de las soluciones que se basan en el uso de Micorservicios y vitualización de contenedores [5]. Por ejemplo, Krylovskiy

Sesbastià Sansó, Universitat de les Illes Balears (UIB), Illes Balears, España, sanso.barcelo94@gmail.com
Carlos Guerrero, Universitat de les Illes Balears (UIB), Illes Balears, España, carlos.guerrero@uib.es *(Corresponding author)
Isaac Lera, Universitat de les Illes Balears (UIB), Illes Balears, España, isaac.lera@uib.es
Carlos Juiz, Universitat de les Illes Balears (UIB), Illes Balears, España, cjuiz@uib.es

et al. [6] propusieron la definición de aplicaciones IoT para Smart City mediante el uso de Microservicios. Pahl y Lee [7] exploraron el uso de contenedores para gestionar las arquitecturas Edge. Un trabajo similar de Ismail et al. [8] utilizó Docker para implementar una plataforma de este tipo.

Probablemente, el trabajo de Bellavista y Zanni [9] es el más cercano a nuestra propuesta, pero con la gran diferencia que está basado en el uso de Docker en lugar de un modelo FaaS. Además, nuestra solución aporta muchas más características en el campo de la interoperabilidad de dispositivos y una mayor sencillez de gestión.

La propuesta de utilizar un modelo basado en FaaS para el despliegue y orquestación de aplicaciones también se ha utilizado en otros entornos anteriormente. Por ejemplo, Roca et al. [10] propusieron este tipo de solución para una arquitectura Fog.

## II. ARQUITECTURA DEL SISTEMA

La arquitectura propuesta soporta la definición de las aplicaciones IoT usando un modelo de FaaS. Esta definición se hace de forma centralizada en un servicio de tipo PaaS, para posteriormente desplegar las funciones necesarias entre los dispositivos IoT. Más concretamente, la arquitectura ha de ser capaz, desde un punto de vista del despliegue de aplicaciones, de: (a) descubrir automáticamente los nuevos dispositivos IoT conectados en el sistema; (b) asociar los dispositivos con las funciones de la aplicación que necesitan ejecutar; y (c) desplegar de forma automatizada las funciones en los dispositivos. Desde un punto de vista del administrador del sistema, la arquitectura ha de ser capaz de: (d) permitir al programador definir funciones que se ejecuten en los dispositivos; (e) permitir la definición de funciones de forma genérica y configurable en cada dispositivo; (f) establecer reglas para la interoperabilidad de los dispositivos.

Nuestra plataforma está dividida en dos capas diferenciadas: la capa de gestión y configuración, tanto de funciones como de dispositivos IoT, y basada en un modelo de PaaS; y la capa de dispositivos IoT, donde se despliegan las aplicaciones IoT basadas en un modelo FaaS, que de forma general, se encargar de los procesos de monitorización o actuación. Dos tipos distintos de comunicación se establecen entre ambas capas: comunicaciones basadas en RESTful, que se utilizan para la interoperabilidad entre las aplicaciones y para el descubrimiento de dispositivos; comunicación basada en el protocolo SSH, que se utiliza para el despliegue de las aplicaciones. La Fig. 1 muestra de forma gráfica todos las capas, componentes y elementos de comunicación del sistema.

El código fuente de la plataforma desarrollada ha sido liberado de forma pública y puede ser descargado y utilizado de forma libre por cualquier usuario[1].

### A. Capa PaaS para la gestión y configuración

Esta capa es el elemento central de la plataforma, que se encarga de gestionar y configurar todo el sistema. Esta capa interacciona, por un lado, con el administrador del sistema para la gestión de las aplicaciones y, por otro lado, con los dispositivos IoT para el despliegue e interoperabilidad. Típicamente se encuentra disponible en un servidor central del entorno IoT, o incluso, en un proveedor de servicios cloud. Su diseño se ha planteado como un entorno de plataforma como servicio (PaaS) de forma que sobre el usuario administrador solo recae la responsabilidad de gestionar el código de las aplicaciones IoT. Para el diseño de esas aplicaciones se ha optado por un patrón de desarrollo basado en definición de funciones, normalmente llamado Function-as-a-Service [11].

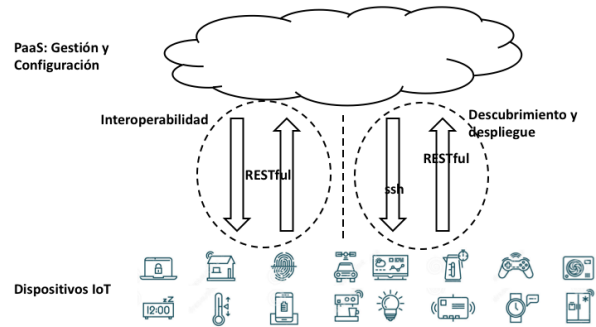

Figura 1. Arquitectura de la plataforma.

La gestión y configuración de las aplicaciones IoT conlleva las siguientes necesidades:

*Gestionar las funciones que forman la aplicación IoT.* Los administradores de la plataforma, o programadores, podrán gestionar las funciones de la aplicación a través de las típicas operaciones CRUD. Además del código fuente del script de la función, también es necesario indicar el intérprete que se ha de utilizar para su ejecución en el dispositivo IoT, indicando el comando que se ha de ejecutar para invocar a la función.

En la Fig. 2 se muestra la pantalla de gestión de funciones, para el caso de una función programada como un script de Python.

*Definir los elementos configurables particulares de cada instancia de una función.* La implementación de las funciones permite incorporar opciones configurables, propias para cada instancia de cada dispositivo.

Un ejemplo ilustrativo es el caso de una función que monitoriza los valores de un sensor. Aunque el código de la función sea el mismo, el puerto a monitorizar donde se encuentra conectado el sensor puede variar. De esta forma, la función que monitoriza los valores podrá ser configurada en el momento del despliegue, bien manualmente por el administrador del sistema, o bien de forma automática con la información proveída desde el dispositivo IoT. Otro ejemplo ilustrativo es el hecho de modificar el tiempo entre envío de los valores monitorizados, ya que estas necesidades pueden variar según el dispositivo o el entorno. La Fig. 2 muestra precisamente una función donde estos dos parámetros han sido incorporados para que puedan ser configurados durante el despliegue.

---

[1] https://github.com/awvaarua/servidor_central

Figura 2. Pantalla de gestión de funciones.

*Definir las reglas de interoperabilidad.* El administrador del sistema podrá definir unas reglas de la forma condición-acción asociadas a un dispositivo, para que se pueda llevar a cabo una interoperabilidad entre los dispositivos del sistema. Los detalles de esta interoperabilidad se incluyen en el apartado D de esta sección. La Fig. 3 muestra la ventana de gestión de las reglas.

*Almacenamiento de los datos monitorizados.* Como una función adicional, nuestra capa de gestión y configuración depende de unos servicios habilitados para que las funciones que se ejecutan en los dispositivos puedan enviar los datos que monitorizan continuamente para que sean almacenados en una base de datos centralizada. De esta forma, se posibilita el posterior análisis de los datos, incluso utilizando técnicas de análisis masivo de datos. El uso de esta funcionalidad simplemente supone llamar periódicamente a un *endpoint* con un conjunto de datos que se han monitorizado durante un periodo de tiempo. Estos pasarán a ser almacenados en la capa centralizada.

Las tecnologías utilizadas para el desarrollo de la capa PaaS de gestión y configuración fueron MongoDB para la persistencia de datos y NodeJS [12] para el desarrollo basado en un patrón MVC [13], y en el que se utilizaron el framework ExpressJS[2] y Passport[3] como sistema de autentificación.

*Definir las reglas de despliegue automatizado.* Cada vez que un nuevo dispositivo IoT sea agregado al sistema, la plataforma lo deberá de identificar y asociar con una serie de funciones a desplegar. Esto lo hará el administrador del sistema de forma manual, pero también se permitirá que se asocien reglas para que mediante las características del nuevo dispositivo, se decida automáticamente las funciones a desplegar. En la versión actual de la plataforma, esta última característica no está totalmente implementada.

### B. Descubrimiento de dispositivos y despliegue de las aplicaciones

Esta parte del sistema es la que establece la forma en que el núcleo central de la plataforma (capa de gestión y configuración) descubre los nuevos dispositivos que se conectan y lleva a cabo el despliegue de las funciones que forman parte del conjunto de la aplicación IoT.

Figura 3. Pantalla de configuración de las reglas de interoperablidad de funciones.

La Fig. 4 muestra el diagrama de secuencia que representa el conjunto de acciones que se llevan a cabo cada vez que se añade un nuevo dispositivo IoT en el sistema. El primer paso que se produce es que el dispositivo debe de notificar a la aplicación de gestión central que se ha añadido a la plataforma, informándole de su dirección IP y de sus características, por ejemplo, sensores que tiene conectado. De esta forma, se podrá realizar el despliegue de las funciones que correspondan, e iniciar la ejecución de las mismas. Por tanto, nada más que arranca el dispositivo, este envía una notificación vía un servicio RESTful al servidor central. Por eso, el dispositivo ha de conocer el *endpoint* (dirección IP del servidor y servicio encargado de añadir el dispositivo).

Una vez que el dispositivo está incluido en el sistema, pueden presentarse dos escenarios. Que el dispositivo conozca y sea capaz de identificar los sensores o actuadores que tiene conectado y que por tanto se pueda llevar a cabo un despliegue automatizado de las aplicaciones que han de ejecutarse en él. Por el contrario, podría ocurrir que el dispositivo no pueda identificar, o no tenga, ningún sensor/actuador. O que no se hayan establecido las reglas automatizadas para desplegar aplicaciones en el caso de los sensores/actuadores que dispone. En este último caso, la aplicación de gestión incluye al dispositivo en una lista de pendientes para que el administrador del sistema decida las aplicaciones que se han de desplegar sobre él.

El despliegue de las aplicaciones se realiza abriendo una conexión SSH entre la aplicación de gestión (que ejerce de cliente) y el dispositivo IoT (que ejerce de servidor), para llevar a cabo la transmisión de las funciones (scripts) que han de ejecutarse en el dispositivo y su posterior ejecución.

---
[2] http://expressjs.com/es/
[3] http://www.passportjs.org/docs/

Gracias a estas decisiones de diseño, se solucionan una serie de problemas relacionados con la heterogeneidad de los dispositivos IoT. Las únicas condiciones que han de satisfacer es que haya disponible alguna implementación de servidor SSH para la arquitectura particular del dispositivo. Dada la gran difusión de este protocolo de comunicación, la mayoría de dispositivos del mercado disponen de versiones servidor del mismo [14].

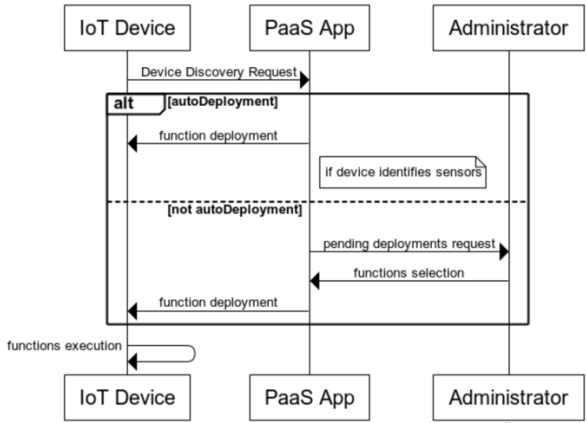

Figura 4. Diagrama de secuencia correspondiente al descubrimiento de dispositivos IoT y despliegue de fucniones.

### C. Capa de dispositivos IoT para el despliegue de aplicaciones FaaS

Esta capa es la que está formada por los dispositivos IoT sobre los que se desplegarán y ejecutarán las funciones de la aplicación IoT. Estos dispositivos incorporan típicamente sensores y actuadores. Uno de los grandes problemas de los sistemas IoT es la gran heterogeneidad en hardware y software de estos dispositivos. Para solucionar este problema se ha optado por llevar a cabo el despliegue y definición de las funciones utilizando estándares de comunicación que están disponibles en un gran número de plataformas, como es por ejemplo el caso de SSH. El resto de procesos de comunicación se basan en peticiones RESTful sobre protocolo HTTP, compatible con cualquier arquitectura.

Para el caso de la ejecución de las funciones, es el programador de la aplicación que deberá de escoger un lenguaje que disponga de intérpretes en todos los dispositivos que formen el ecosistema IoT para el que esté desarrollando su aplicación.

Así que, de forma resumida, los dispositivos IoT deberán cumplir dos condiciones: disponer de un servidor SSH y disponer de un intérprete para el código en el que están escritas las funciones de la aplicación.

Finalmente, estos dispositivos deberán de configurarse para que en el momento de iniciarse por primera vez, realicen una petición RESTful al *endpoint* de nuestra plataforma que se encarga del descubrimiento de los nuevos dispositivos. Para ello es suficiente incluir un pequeño script que se ejecute al iniciar el dispositivo.

### D. Interoperabilidad de las aplicaciones y dispositivos

Adicionalmente de la interoperabilidad que el programador de aplicación pueda implementar en las propias funciones, la plataforma posibilita al administrador la definición de ciertas reglas básicas para de la forma condición-acción. De esta forma, el administrador del sistema podrá hacer que un dispositivo que incorpore un actuador lleve a cabo una acción cuando se cumpla alguna condición sobre los valores monitorizados sobre otros dispositivos. Como ejemplo simple, nos podemos imaginar una vivienda que incorpore un dispositivo con un sensor de temperatura y otro dispositivo con un relé para que active la calefacción bajo ciertos valores de temperatura. La Fig. 3 muestra la pantalla del sistema correspondiente con un conjunto de reglas muy sencillas. En ella, se inicia la grabación de un video de 5 segundos en un nodo A y un nodo B, cuando el nodo A detecta un movimiento en un sensor de movimiento.

Como característica adicional, la plataforma también permite añadir una acción de tipo especial que es la de enviar un mensaje a un usuario de Telegram a través de un *bot* que se desarrolló con ese objetivo.

## III. EVALUACIÓN Y RESULTADOS

### A. Entorno de Pruebas

Existen muchos dominios de aplicación para un ecosistema IoT [15]. En nuestro caso de estudio, hemos escogido un ejemplo basado en una vivienda inteligente (*Smart Homes*). Hemos implementado un pequeño *testbed* de laboratorio para un sistema de vigilancia de cámaras, con sensores de movimiento que detectan algún tipo de actividad. Cuando se produce este hecho, las cámaras del sistema se encargan de grabar un video de la escena. Durante la grabación, la iluminación de la zona se activa mediante una bombilla que se encuentra conectada a un relé. El video es almacenado en el servidor central, y también es enviado por el bot de Telegram al teléfono del propietario de la vivienda. Para tener más claro el funcionamiento del sistema, existe un video donde se puede ver un ejemplo del mismo[4].

Todo el sistema ha sido implementado utilizando Raspberry Pi 2 model B. La Tabla I muestra la función y los componentes de las 3 placas utilizadas.

TABLA I. DISPOSITIVOS DEL TESTBED PARA SMART-HOME.

| ID | FUNCIÓN | COMPONENTES |
|---|---|---|
| RB1 | DISPOSITIVO IOT | SENSOR MOVIMIENTO (PIR FROTAL) |
| RB2 | DISPOSITIVO IOT | SENSOR CÁMARA (PI CAMERA) ACTUADOR RELÉ BOMBILLA |
| RB3 | SERVIDOR CLOUD | --- |

Los scripts utilizados para implementar la aplicación IoT del *testbed* se encuentran disponibles en línea[5]. Todos ellos han sido desarrollados con Python 2.7. Las placas Raspberry que actuaban de dispositivos IoT fueron instaladas con una imagen modificada del sistema operativo Raspbian, en su versión Jessie, propio de las arquitecturas Raspberry Pi. A la imagen base del SO, se le incluyeron los scripts Python de nuestro repositorio[6], y se configuró el archivo */etc/rc.local* para que ejecutara el script *init.py* en el arranque. Dicho script es el que se encarga de realizar la petición RESTful encargada de iniciar el proceso de descubrimiento del dispositivo. Dicha imagen modificada también ha sido colgada on-line[7].

---

[4] https://youtu.be/ 3bpwIy2xG9A
[5] https://github.com/awvaarua/monitorizacion

Para el caso de la placa Raspberry que ejercía las funciones de servidor centralizado, se instaló igualmente un sistema operativo Raspbian Jessie y los paquetes *nodejs* y *mongodb-server*, necesarios para ejecutar nuestra aplicación de gestión y configuración del sistema IoT. La aplicación se instaló desde los repositorios que ya se han comentado que contienen todo nuestro sistema de gestión.

La Fig. 5 muestra el detalle de algunos de los dispositivos que forman parte del sistema.

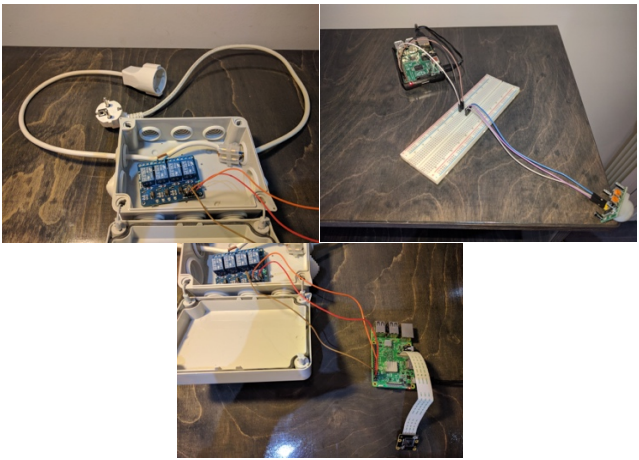

Figura 5. *Testbed* utilizado para el caso de estudio de nuestra plataforma.

*A. Resultados*

Las dos principales ventajas de nuestra plataforma son la sencillez de despliegue de la aplicación IoT en los dispositivos, y la reducción de los recursos consumidos en dichos dispositivos en frente de otras alternativas, como por ejemplo sería el uso de contenedores (*containers* de Docker) para desplegar las aplicaciones.

En este apartado presentamos un análisis de la segunda de las mayores ventajas de nuestra solución. Para ello hemos comparado el uso de recursos que se genera con el caso de ejemplo presentado cuando utilizamos nuestra plataforma, y cuando se usa una plataforma basada en Docker, como por ejemplo la que se plantea en [9]. Para ello, hemos medido el uso de recursos (CPU y memoria principal) en el sistema a medida que se incrementa el número de funciones (o *containers* en el caso con el que comparamos). Para ello hemos utilizado el script de la función que monitoriza el sensor de movimiento. Las Figs. 6 Y 7 muestran la comparación entre ambas alternativas: nuestra plataforma etiquetada como *FaaS*; y la opción con la que comparamos, etiquetada como *Docker*, en la que se utilizó un contenedor con la imagen de Python 2.7 incluyendo el script comentado.

---

[6] https://github.com/awvaarua/nodo
[7] https://github.com/awvaarua/raspbian_link

Se puede observar como en ambos casos, CPU y memoria principal, nuestra alternativa consume bastante menos recursos. Es interesante destacar que para el caso de Docker solo se muestran los valores hasta los 60 contenedores, ya que la placa Raspberry Pi 2 modelo B no consiguió desplegar un número de contenedores superior, y el sistema acababa cayendo. Tras el análisis de los números, probablemente este hecho se producía debido a que la memoria principal se agotaba.

También medimos los consumos de recursos de los servicios necesarios para implantar cada una de las soluciones. En el caso de nuestra plataforma fue necesario activar el servicio de SSH, que suponía un consumo de 0.3% de CPU y 1,5% de memoria. Por el contrario, los servicios necesarios para Docker, consumían 1,4% de CPU y 5,3% de memoria. De nuevo se comprueba que nuestra solución supone una reducción en los recursos consumidos.

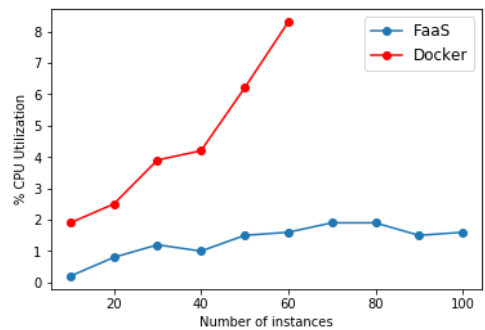

Figura 6. Comparativa del uso de CPU entre nuestra plataforma y una solución basada en Docker.

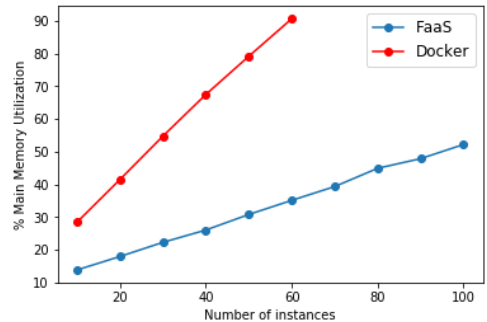

Figura 7. Comparativa del uso de memoria principal entre nuestra plataforma y una solución basada en Docker.

TABLA II. COMPARACIÓN DEPSLIEGUE APLICACIONES.

| MÉTRICA    | FAAS    | DOCKER        |
|------------|---------|---------------|
| TAMAÑO     | 1.0 KB  | 350 MB        |
| TIEMPO (S) | 13 SEG. | 4 MIN. 23 SEG.|

Es interesante remarcar también, otras métricas, como los tiempos y tamaños de transferencia necesarios para desplegar las funciones, o *containers*, en cada nodo (Tabla II). En nuestro caso fue suficiente con transmitir aproximadamente 1 KB del tamaño del script de la función. En el caso de la solución con Docker, durante el despliegue de la primera función, el servicio de Docker se descargó una imagen con Python que ocupaba aproximadamente 350 MB durante unos 4 minutos.

## VII. CONCLUSIÓN

Este artículo presenta una plataforma que reduce la complejidad y el uso de recursos del despliegue de aplicaciones IoT usando un patrón *Function-as-a-Service*. A parte de un despliegue autónomo de las funcioens, nuestra plataforma también ofrece la posibilidad de definir reglas de interoperabilidad entre dispositivos, a parte de las que puedan establecer los programadores dentro del código de las propias funciones. Finalmente, también existe la posibilidad de almacenar datos obtenidos de los sensores del sistema para que sean procesados posteriormente.

Se ha llevado a cabo una evaluación de nuestra propuesta utilizando como caso de ejemplo un ambiente de Smart Home, donde se ha desplegado una aplicación de grabación de video controlado por un sensor de movimiento. Se ha visto que la sobrecarga generada por nuestra plataforma es muy reducida, y que los tiempos de despliegue y gestión también se reducen en comparación a otras alternativas.

Como trabajos futuros, en primer lugar, se plantea la posibilidad de extender esta plataforma a una arquitectura Fog, En ellas, las aplicaciones IoT pueden incorporar funciones que por su necesidad de recursos son ejecutadas en un servidor central (típicamente un servicio cloud) en lugar de en los dispositivos IoT. Para mejorar los aspectos de latencia, las arquitecturas Fog proponen ejecutar dichas funciones en los nodos intermedios de la red de comunicación. Nuestra plataforma podría extenderse para que se pueda llevar a cabo un despliegue de funciones también en nodos de comunicación y no solo en los dispositivos IoT.

Un segundo trabajo futuro sería agregar a la herramienta la posibilidad de definir funciones también con lenguajes compilados y no únicamente interpretados como ocurre actualmente. De esta forma, la plataforma, cada vez que desplegara una función en un dispositivo, debería de reconocer la arquitectura del mismo y compilar un ejecutable para esa arquitectura, para finalmente desplegar el ejecutable en el dispositivo.

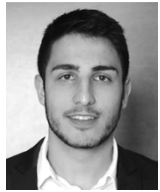

**Sebastià Sansó** received his Bachelor degree in computer engineering at the Balearic Islands University in 2017. He is currently a computer programmer in an international enterprise and he collaborates in some research activities with the University of Balearic Islands. His main interests are Internet of Things and web programming.

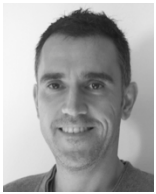

**Carlos Guerrero** received his Ph.D. degree in Computer Engineering at the Balearic Islands University in 2012. He is an assistant professor of Computer Architecture and Technology at the Computer Science Department of the University of the Balearic Islands. His research interests include web performance, performance optimization, resource management, web engineering, cloud computing, fog computing, and IoT. He has authored around 40 papers in different international conferences and journals.

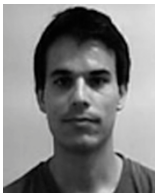

**Isaac Lera** received his Ph.D. degree in Computer Engineering at the Balearic Islands University in 2012. He is an assistant professor of Computer Architecture and Technology at the Computer Science Department of the University of the Balearic Islands. His research lines are semantic web, open data, system performance, educational innovation and human mobility. He has authored in several journals and international conferences.

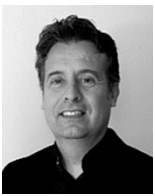

**Carlos Juiz** received his Ph.D. degree in Computer Engineering at the Balearic Islands University in 2001. He is an associate professor of Computer Architecture and Technology at the Computer Science Department of the University of the Balearic Islands. His research interests include performance engineering, cloud comput- ing and IT governance. He has authored around 150 papers in different international conferences and journals.